\documentclass[
a4paper,12pt]{article}
\usepackage{amsmath}
\usepackage{amsfonts}
\usepackage{amssymb}
\usepackage{hyperref}
\usepackage{cite}

\def\be{\begin{equation}}
\def\ee{\end{equation}}
\def\arr{\begin{array}{rll}}
\def\ea{\end{array}}
\def\bea{\begin{eqnarray}}
\def\eea{\end{eqnarray}}

\renewcommand{\thefootnote}{\arabic{footnote}}
0

\begin{document}
\renewcommand{\thefootnote}{\fnsymbol{footnote}}
\vskip 1cm
\begin{center}
{\LARGE\bf  Remark on near-horizon geometry  }
\vskip 0.5cm
{\LARGE\bf of extreme regular black holes }\\
\vskip 1cm
$
\textrm{\Large Sergei Filyukov }
$
\vskip 0.7cm
{\it
Tomsk Polytechnic University, 634050 Tomsk, Lenin Ave. 30, Russia} \\
\vskip 0.4cm
{E-mail: filserge@tpu.ru}

\end{center}
\begin{abstract} \noindent
It is shown that the near horizon geometry of a generic extreme regular black hole solution of Einstein gravity coupled to nonlinear electrodynamics is described by the $AdS_2 \times S^2$ spacetime.
\end{abstract}

\vskip 1cm

The first attempt to rigorously construct a non-singular black hole metric dates back to the work of Bardeen \cite{Bar} in which the singularity was replaced by the de Sitter core. Although the ensuing
metric did not solve the vacuum Einstein equations, it was later realized how to consistently incorporate it into an exact solution of nonlinear electrodynamics coupled to gravity \cite{Garcia}.
Since then a variety of regular black hole metrics have been constructed both in the context of nonlinear electrodynamics coupled to gravity and alternative theories of gravity \cite{Beato3}--\cite{Manuel1}.

As is well known, in the near horizon region the isometry group of the conventional {\it extreme} black hole metrics is enhanced to include the conformal factor $SO(2,1)$ \cite{Bardeen,KL}. This conformal symmetry paved the way for numerous applications in the context of (super)conformal mechanics \cite{Kallosh}--\cite{AG1} and (super)integrable systems \cite{BNY}--\cite{AS}. It is then natural to wonder what is the near horizon limit of the {\it extreme} regular black hole. The goal of this letter is to study this issue. Below we first describe a generic extreme regular black hole solution of Einstein gravity coupled to nonlinear electrodynamics and then demonstrate that in the near horizon limit it reduces the $AdS_2 \times S^2$ spacetime.

Our starting point is a generic static metric
\be\label{metric}
ds^2=-f(r)dt^2+\frac{dr^2}{f(r)}+r^2 d\Omega^2_2,
\ee
where it is assumed that $f(0)=1$ and $f'(0)=0$, for the metric to be non-singular at the origin.
For the extreme case the inner and outer horizons coalesce and $f(r)$ has double zero at the horizon radius $r_h$
\be\label{f}
f(r)=a(r) (r-r_h)^2,
\ee
where $a(r)$ is nonzero at $r_h$, $a(0)=\frac{1}{r_h^2}$ and $a'(0)=\frac{2}{r_h^3}$ but otherwise arbitrary. The latter conditions guarantee nonsingular behaviour of curvature invariants thus yielding a regular extreme black hole metric.

As the next step, consider the action functional of the Einstein gravity coupled to nonlinear electrodynamics
\be
S=-\frac{1}{16 \pi}\int d^4 x \sqrt{|g|} \left(R+2 \mathcal L(F)\right)\label{Act},
\ee
where $F=\frac{1}{4} F_{\mu \nu} F^{\mu \nu}$ and $\mathcal L(F)$ is a function to be fixed below. The corresponding equations of motion read
\bea
&&
R_{\mu \nu}-\frac{1}{2}R g_{\mu \nu}= \left( \mathcal L g_{\mu \nu}+\mathcal L' F_{\mu \alpha}F^\alpha_{ ~\nu} \right), \qquad \partial_\mu (\sqrt{|g|}\mathcal L' F^{\mu \nu})=0\label{Ein},
\eea
where $\mathcal L'=\frac{d \mathcal L}{d F}$.

Let us first focus on the purely magnetic case characterized by the only non--vanishing component of the field strength tensor $F_{\theta\phi}=P \sin{\theta}$, where $P$ is a magnetic charge. This yields
\bea
F=\frac{P^2}{2 r^4}\label{sqrm} \quad \Rightarrow \quad
\frac{d}{d F}=\frac{dr}{dF}\frac{d}{dr}=-\frac{r^5}{2 P^2}\frac{d}{dr}\label{der}
\eea
and reduces the Einstein equations to the condition which links $\mathcal L(F)$ to $f(r)$
\bea
\mathcal L=\frac{r f'+f-1}{r^2}\label{lagr},\label{ad}
\eea
Thus for each $f(r)$ one has a magnetic monopole coupled to gravity described by the action functional (\ref{Act}) with $\mathcal L$ in (\ref{ad}) and $F$ related to $r$ via (\ref{der}). By construction the solution is regular. A peculiar feature of the magnetic solution is that (\ref{lagr}) and~(\ref{sqrm}) tend to constant values when implementing the near horizon limit
\bea
\mathcal L  \to  -\frac{1}{r_h^2}, \qquad
F  \to  \frac{P^2}{2 r_h^4}.
\eea

An alternative possibility is to consider an electric charge $Q$ and the field strength
\be
F_{tr}=\frac{Q}{r^2 \mathcal L'}.
\ee
It reduces the Einstein equations to
\bea
\mathcal L=\frac{1}{2 }(f''+2\frac{f'}{r})\label{electro}
\eea
and allows one to express $F$ as a function of $r$
\be
F=\frac{-Q^2}{2 r^4 \mathcal L'^2}=-\frac{(r^2 f''-2f+2)^2}{8  Q^2},\label{sqre}
\ee
which in the near horizon limit tends to
\bea
F \to \frac{(r_h a(r_h)+1)^2}{2 Q^2}.
\eea
One can also introduce the magnetic charge $P$
\be
F_{\mu\nu}dx^\mu\wedge dx^\nu=\frac{Q}{r^2 \mathcal L'}dt\wedge dr+P \sin\theta d\theta\wedge d\phi, \quad \Rightarrow \quad F=\frac{1}{2 r^4}(P^2-\frac{Q^2}{\mathcal L'^2})
\ee
The Einstein equations give
\bea
\mathcal L+\frac{Q^2}{r^4 \mathcal L'}=\frac{r f'+f-1}{r^2}\label{Glagr},\qquad
\mathcal L-\mathcal L'\frac{P^2}{r^4}=\frac{1}{2}(f''+2 \frac{f'}{r})\label{Gad},
\eea
which allow one to fix $\mathcal L$ and $\mathcal L'$ algebraically.

To summarize, by appropriately choosing $\mathcal L$ in (\ref{Act}) one can always  promote a regular extreme black hole metric (\ref{metric}), (\ref{f}) to an exact solution of the Einstein gravity coupled to nonlinear electrodynamics (\ref{Ein}).

At this point the near horizon limit can be implemented along the lines in \cite{Bardeen,KL}. One first changes the coordinates 
\be
r\rightarrow r_h+\epsilon r, \quad t\rightarrow \frac{t}{\epsilon}, 
\ee
and then sends $\epsilon$ to zero. The first redefinition ensures that, as $\epsilon\rightarrow 0$, one approaches the horizon from the outer region. The subsequent rescaling of the temporal variable is needed so as to guarantee that the resulting metric is nonsingular as $\epsilon\rightarrow 0$. For the regular black hole metric (\ref{metric}) this yields
\be\label{ans}
ds^2=-a(r_h)r^2 dt^2+\frac{dr^2}{a(r_h)r^2}+r_h^2 d\Omega^2_2.
\ee
with $a(r)$ defined in (\ref{f}).

As a result, the near horizon metric coincides with $AdS_2 \times S^2$ in Poincar\'e coordinates with the $SO(2,1)\times SO(3)$ isometry group. It may seem puzzling that starting with a regular black hole solution, one finally obtains a metric which has a coordinate singularity at $r=0$. The reason is that the Poincar\'e coordinates cover only half of the spacetime. In order to remove the coordinate singularity and construct a metric in global coordinates, it suffices to implement the transformation
\be
r=\sqrt{1+y^2} \cos{\tau} +y, \qquad t=\frac{\sqrt{1+y^2} \sin{\tau}}{a(r_h) r},
\ee
where $y$ and $\tau$ are new radial and temporal variables, 
which yields the near horizon metric of a regular black hole in global coordinates
\be
ds^2=\frac{1}{a(r_h)}\left(-(1+y^2) d\tau^2+\frac{dy^2}{1+y^2}\right)+r_h^2 d\Omega^2_2.
\ee
Along with the near horizon value of the field strength $F_{\mu\nu}dx^\mu\wedge dx^\nu$ it reproduces the Bertotti-Robinson solution of the Einstein--Maxwell equations.

We thus conclude that, irrespective of the details of a concrete regular extreme black hole configuration, its near horizon geometry is universal and described by $AdS_2 \times S^2$.

\vspace{0.5cm}

\noindent{\bf Acknowledgements}\\

\noindent
This work was supported by the Tomsk Polytechnic University competitiveness enhancement program and RFBR grant 18-52-05002.




\begin{thebibliography}{99}
\bibitem{Bar}
J.M. Bardeen, \textit{Non-singular general relativistic gravitational collapse}, Proceedings of the International Conference GR5, Tbilisi, U.S.S.R., 1968.
\bibitem{Garcia}
 E. Ayon-Beato, A. Garcia, \textit{The Bardeen model as a nonlinear magnetic monopole},
Phys. Lett. B {\bf 493} (2000) 149, gr-qc/0009077.
\bibitem{Beato3}
E. Ayon-Beato, A. Garcia, \textit{Regular black hole in general relativity coupled to nonlinear electrodynamics},
Phys. Rev. Lett. {\bf 80} (1998) 5056, gr-qc/9911046.
\bibitem{Beato2}
E. Ayon-Beato, A. Garcia, \textit{New regular black hole solution from nonlinear electrodynamics},
Phys. Lett. B {\bf 464} (1999) 25, hep-th/9911174.
\bibitem{Beato4} E. Ayon-Beato, A. Garcia,
\textit{Four parametric regular black hole solution},
Gen. Rel. Grav. {\bf 37} (2005) 635, arXiv:hep-th/0403229.
\bibitem{Kirill1}
K.A. Bronnikov,
\textit{Regular magnetic black holes and monopoles from nonlinear electrodynamics},
Phys. Rev. D {\bf 63} (2001) 044005, gr-qc/0006014.
\bibitem{Kirill12}
K.A. Bronnikov, \textit{Comment on "Regular black hole in general relativity coupled to nonlinear electrodynamics"},
Phys. Rev. Lett. {\bf 85} (2000) 4641.



\bibitem{Irina}
I. Dymnikova, \textit{Regular electrically charged structures in nonlinear electrodynamics coupled to general relativity},
Class. Quant. Grav. {\bf 21} (2004) 4417, gr-qc/0407072.

\bibitem{Reff}
Z. Fan, X. Wang,
\textit{Construction of Regular Black Holes in General Relativity},
Phys. Rev. D {\bf 94} (2016) 124027, gr-qc/1610.02636.

\bibitem{berej} W. Berej, J. Matyjasek, D. Tryniecki, M. Woronowicz,
\textit{Regular black holes in quadratic gravity},
Gen. Rel. Grav. {\bf 38} (2006) 885,  arXiv:hep-th/0606185.
\bibitem{Stefano} S. Ansoldi, \textit{Spherical black holes with regular center: a review of existing models including a recent realization with Gaussian sources},
arXiv:0802.0330.
\bibitem{Nami} N. Uchikata, S. Yoshida, T. Futamase, \textit{New solutions of charged regular black holes and their stability},
Phys. Rev. D {\bf 86} (2012) 084025, arXiv:1209.3567.


\bibitem{Rot1} C. Bambi, L. Modesto,
\textit{Rotating regular black holes},
Phys. Lett. B {\bf 721} (2013) 329, arXiv:1302.6075.
\bibitem{balart}
L. Balart, E.C. Vagenas, \textit{Regular black hole metrics and the weak energy condition},
Phys. Lett. B {\bf 730} (2014) 14, arXiv:1401.2136.
\bibitem{Rot2} J.C.S. Neves, A. Saa,
\textit{Regular rotating black holes and the weak energy condition},
Phys. Lett. B {\bf 734} (2014) 44, arXiv:1402.2694.
\bibitem{Rot3} B. Toshmatov, B. Ahmedov, A. Abdujabbarov, Z. Stuchlik,
\textit{Rotating regular black hole solution},
Phys. Rev. D {\bf 89} (2014) 104017, arXiv:1404.6443.
\bibitem{Rot4} M. Azreg-Ainou,
\textit{Generating rotating regular black hole solutions without complexification},
Phys. Rev. D {\bf 90} (2014) 064041, arXiv:1405.2569.
\bibitem{Leonardo2}
L. Balart, E.C. Vagenas, \textit{Regular black holes with a nonlinear electrodynamics source},
Phys. Rev. D {\bf 90} (2014) 124045, arXiv:1408.0306.
\bibitem{Rot5} S. G. Ghosh, S. D. Maharaj,
\textit{Radiating Kerr-like regular black hole},
Eur. Phys. J. C {\bf 75} (2015) 7, arXiv:1410.4043.
\bibitem{rodrigues2} M.E. Rodrigues, E.L.B. Junior, G.T. Marques, V.T. Zanchin,
\textit{Regular black holes in $f(R)$ gravity coupled to nonlinear electrodynamics},
Phys. Rev. D {\bf 94} (2016) 024062, arXiv:1511.00569.
\bibitem{Frol} V.P. Frolov,
\textit{Notes on nonsingular models of black holes},
Phys. Rev. D {\bf 94} (2016) 104056, arXiv:1609.01758.
\bibitem{rodrigues1} E.L.B. Junior, M.E. Rodrigues, M.J.S. Houndjo,
\textit{Regular black holes in $f(T)$ Gravity through a nonlinear electrodynamics source},
JCAP {\bf 1510} (2015) 060, arXiv:1503.07857.
\bibitem{Rot6} I. Dymnikova, E. Galaktionov,
\textit{Regular rotating electrically charged black holes and solitons in non-linear electrodynamics minimally coupled to gravity},
Class. Quant. Grav. {\bf 32} (2015) 165015, arXiv:1510.01353.
\bibitem{Rot7} T. De Lorenzo, A. Giusti, S. Speziale,
\textit{Non-singular rotating black hole with a time delay in the center},
Gen. Rel. Grav. {\bf 48} (2016) 31, arXiv:1510.08828.
\bibitem{rodrigues3} M.E. Rodrigues, J.C. Fabris, E.L.B. Junior, G.T. Marques,
\textit{Generalisation for regular black holes on general relativity to $f(R)$ gravity},
Eur. Phys. J. C {\bf 76} (2016)  250,  arXiv:1601.00471.
\bibitem{Rot8} R. Torres, F. Fayos,
\textit{On regular rotating black holes},
Gen. Rel. Grav. {\bf 49} (2017) 2, arXiv:1611.03654.

\bibitem{CuletuR} H.~Culetu,
\textit{On a regular charged black hole with a nonlinear electric source},
Int.\ J.\ Theor.\ Phys. {\bf 54} (2015), arXiv:1408.3334.
\bibitem{Ponce} J. Ponce de Leon,
\textit{Regular Reissner-Nordstr\"om black hole solutions from linear electrodynamics},
Phys. Rev. D {\bf 95} (2017) 124015, arXiv:1706.03454.

\bibitem{Culetu} H.~Culetu,
\textit{On a regular modified Schwarzschild spacetime },
arXiv:1305.5964.

\bibitem{Nicolini} P.~Nicolini,
\textit{Noncommutative Black Holes, The Final Appeal To Quantum Gravity: A Revie}, Int.\ J.\ Mod.\ Phys.\ A {\bf 24}, 1229 (2009)
arXiv:0807.1939.




\bibitem{Manuel1} M.E. Rodrigues, E.L.B. Junior, M.V. de S. Silva,
\textit{Using dominant and weak energy conditions for build new classes of regular black holes}, JCAP {\bf 02} (2018) 059,
arXiv:1705.05744.
\bibitem{Bardeen}  J.M. Bardeen, G.T. Horowitz,
\textit{The extreme Kerr throat geometry: a vacuum analog of $AdS_2 \times S^2$},
Phys. Rev. D {\bf 60} (1999) 104030,  arXiv:9905099.
\bibitem{KL}
H.K. Kunduri, J. Lucietti, \textit{Classification of near-horizon geometries of extremal black holes}, Living Rev. Rel. {\bf 16}
(2013) 8, arXiv:1306.2517.
\bibitem{Kallosh}
P. Claus, M. Derix, R. Kallosh, J. Kumar, P.K. Townsend, A. Van Proeyen, {\it Black holes and superconformal mechanics}, Phys. Rev. Lett. {\bf 81} (1998) 4553, hep-th/9804177.
\bibitem{Townsend}
J.A. de Azcarraga, J.M. Izquierdo, J.C. Perez Bueno, P.K. Townsend, {\it Superconformal mechanics and nonlinear realizations}, Phys. Rev. D {\bf 59} (1999) 084015, arXiv:hep-th/9810230.
\bibitem{BIK3}
S. Bellucci, E. Ivanov, S. Krivonos, {\it AdS/CFT equivalence transformation}, Phys. Rev. D {\bf 66} (2002) 086001, hep-th/0206126.
\bibitem{ikn}
E. Ivanov, S. Krivonos, J. Niederle, {\it Conformal and superconformal mechanics revisited}, Nucl. Phys. B {\bf 677} (2004) 485, hep-th/0210196.
\bibitem{bgik}
S. Bellucci, A. Galajinsky, E. Ivanov, S. Krivonos, {\it  	
${AdS}_2/{CFT}_1$, canonical transformations and superconformal mechanics}, Phys. Lett. B {\bf 555} (2003) 99, hep-th/0212204.
\bibitem{LP}
C. Leiva, M. Plyushchay, {\it Conformal symmetry of relativistic and nonrelativistic systems and $Ads/CFT$ correspondence}, Annals Phys. {\bf 307} (2003) 372,
hep-th/0301244.
\bibitem{Anabalon_Zanelli}
A. Anabalon, J. Gomis, K. Kamimura, J. Zanelli, {\it $N=4$ superconformal mechanics as a non--linear realization}, JHEP 0610 (2006) 068, hep-th/0607124.
\bibitem{g1}
A. Galajinsky, {\it  	
Particle dynamics on ${AdS}_2 \times {S}^2$ background with two-form flux}, Phys. Rev. D {\bf 78} (2008) 044014, arXiv:0806.1629.
\bibitem{G2}
A. Galajinsky, {\it Particle dynamics near extreme Kerr throat and supersymmetry}, JHEP {\bf 1011} (2010) 126, arXiv:1009.2341.
\bibitem{GO}
A. Galajinsky, K. Orekhov, {\it N=2 superparticle near horizon of extreme Kerr-Newman-AdS-dS black hole}, Nucl. Phys. B {\bf 850} (2011) 339, arXiv:1103.1047.
\bibitem{BelK}
S. Bellucci, S. Krivonos, {\it $N=2$ supersymmetric particle near extreme Kerr throat}, JHEP {\bf 1110} (2011) 014, arXiv:1106.4453.
\bibitem{gn}
A. Galajinsky, A. Nersessian, {\it  	
Conformal mechanics inspired by extremal black holes in $d=4$ }, JHEP {\bf 1111} (2011) 135, arXiv:1108.3394.
\bibitem{BSW}
A. Galajinsky, {\it Particle collisions on near horizon extremal Kerr background}, Phys. Rev. D {\bf 88} (2013) 027505, arXiv:1301.1159.
\bibitem{Gal}
A. Galajinsky, {\it $N=4$ superconformal mechanics from the $su(2)$ perspective},  JHEP {\bf 1502} (2015) 091, arXiv:1412.4467.
\bibitem{AG1}
A. Galajinsky, \textit{ Couplings in $D(2,1;\alpha)$ superconformal mechanics from the $SU(2)$ perspective}, JHEP {\bf 1703} (2017) 054, arXiv:1702.01955.
\bibitem{BNY}
S. Bellucci, A. Nersessian, V. Yeghikyan, {\it Action--angle variables for the particle near extreme Kerr throat},  Mod. Phys. Lett. A {\bf 27} (2012) 1250191, arXiv:1112.4713.
\bibitem{Sag}
A. Saghatelian, {\it Near--horizon dynamics of particle in extreme Reissner-Nordstr\"om and Clement-Gal'tsov black hole backgrounds: action-angle variables}, Class. Quant. Grav. {\bf 29} (2012) 245018, arXiv:1205.6270.
\bibitem{Gal1}
A. Galajinsky, {\it Near horizon black holes in diverse dimensions and integrable models}, Phys. Rev. D {\bf 87} (2013) 024023, arXiv:1209.5034.
\bibitem{GNS}
A. Galajinsky, A. Nersessian, A. Saghatelian, {\it Superintegrable models related to near horizon extremal Myers-Perry black hole in arbitrary dimension},  JHEP {\bf 1306} (2013) 002, arXiv:1303.4901.
\bibitem{GNS1}
A. Galajinsky, A. Nersessian, A. Saghatelian, {\it Action--angle variables for spherical mechanics related to near horizon extremal Myers-–Perry black hole}, J. Phys. Conf. Ser. {\bf 474} (2013) 012019.
\bibitem{HNS}
T. Hakobyan, A. Nersessian, M.M. Sheikh--Jabbari, {\it Near horizon extremal Myers-Perry black holes and integrability of associated conformal mechanics}, Phys. Lett. B {\bf 772} (2017) 586, arXiv:1703.00713.
\bibitem{HD}
H. Demirchian, {\it Note on constants of motion in conformal mechanics associated with near horizon extremal Myers--Perry black holes},  Mod. Phys. Lett. A {\bf 32} (2017) 1750144, arXiv:1706.04861.
\bibitem{AS}
H. Demirchian, A. Nersessian, S. Sadeghian, M.M. Sheikh-Jabbari, {\it Integrability of geodesics in near-horizon extremal geometries: Case of Myers-Perry black holes in arbitrary dimensions}
Phys. Rev. D {\bf 97} (2018) 104004, arXiv:1802.03551.


\end{thebibliography}
\end{document}